\documentclass[twocolumn,showkeys,showpacs,preprintnumbers,amsmath,amssymb]{revtex4}

\usepackage{graphicx}% Include figure files
\usepackage{dcolumn}% Align table columns on decimal point
\usepackage{bm}% bold math

\begin{document}

\preprint{ver. 2}

\title{Spin transfer switching of spin valve nanopillars using nanosecond pulsed currents\footnote{Contribution of NIST, an agency of the U.S. government, not subject to copyright.}}% Force line breaks with \\

\author{Shehzaad Kaka}
\email{kaka@boulder.nist.gov, fax: (303)497-7364}
%\homepage{Publication of NIST, not subject to copyright}
\author{Matthew R. Pufall}%
 \author{William H. Rippard}
 \author{Thomas J. Silva}
 \author{Stephen E. Russek}
 \affiliation{National Institute of Standards and Technology, 325 Broadway Boulder, CO 80305}%

\author{Jordan A. Katine}
\author{Matthew Carey}
\affiliation{Hitachi San Jose Research Center, 650 Harry Road, San Jose, CA 95120}%

\date{\today}% It is always \today, today,

\begin{abstract}
Spin valve nanopillars are reversed via the mechanism of spin momentum transfer using current pulses applied perpendicular to the film plane of the device.  The applied pulses were varied in amplitude from 1.8 mA to 7.8 mA, and varied in duration within the range of 100 ps to 200 ns. The probability of device reversal is measured as a function of the pulse duration for each pulse amplitude.  The reciprocal pulse duration required for 95\,\% reversal probability is linearly related to the pulse current amplitude for currents exceeding 1.9 mA.  For this device, 1.9 mA marks the crossover between dynamic reversal at larger currents and reversal by thermal activation for smaller currents.

%boundary whereby greater currents drive fully dynamic reversal and lesser currents cause reversal by thermal activation. 
\end{abstract}

%Use showkeys class option if keyword
\maketitle

\section{\label{sec:intro}Introduction}

The free-layer magnetization in a spin valve can be reversed by a sufficiently high current density applied perpendicular to the film plane\cite{albert1sw,albert2sw,urazhdin_smt}.  This effect, explained by angular momentum transfer between spin polarized currents and the magnetization of the ferromagnetic films within the spin valve structure\cite{slonc1}, is being explored as a possible high-speed bit-writing mechanism for magnetic random access memory (MRAM).  Such a mechanism may replace conventional field-driven bit reversal within the cross-point writing scheme\cite{Tehrani2000} by eliminating the problem of possible reversal of half-selected bits  in ultra-dense memory arrays.  
%(In addition, direct current induced reversal in very small devices may require less power than reversal due to in plane fields %generated from currents in wires near the device(do calc)).  
In previous work, current-driven switching has been experimentally investigated with currents varying from quasi-DC, to millisecond-duration ramps, to pulse widths down to 100 ns\cite{albert2sw,myers1sw,koch1swcpp}.  In this paper we investigate the reversal of current-perpendicular-to-plane (CPP) spin valve nanopillars due to pulsed currents whose durations range from 200 ns to 100 ps.

\section{\label{sec:experiment}Experiment}

Devices having lateral dimensions less than 200 nm are required to achieve spin transfer switching because large current densities ($\approx 10^7$ A/cm$^2$) are needed.  We have studied 50 nm $\times$ 100 nm 
spin valve nanopillars sputter-deposited with the following layer composition (from bottom to top): IrMn(7)/Co(7.5)/Cu(4)/Co(3).  Here the layer thicknesses in parentheses are given in nanometers.   The top Co layer is referred to as the free-layer because its smaller thickness gives it both a lower coercivity and lower critical current. The top and bottom of the spin valve structure are in direct contact with thick metallic leads that connect to contact pads on the chip surface.  The current flows perpendicular to the film plane, and, for the purpose of this report, positive current specifies electron flow from the top lead to the bottom lead. 

The nanopillar structure is patterned by electron beam lithography and further fabricated through a process similar to what has been previously described\cite{katine1cpp}.  The device shape is an elongated hexagon that provides a uniaxial shape anisotropy.  A quasi-static resistance vs. magnetic field hysteresis loop for the device studied in this paper is shown in Fig. 1(a).  The resistance change between the antiparallel state and the parallel state is about 200 m$\Omega$, with abrupt switching between states.  The magnetic layers of the device are antiparallel in zero applied field because of the magnetostatic interaction between effective magnetic charges at the edges of the layers.   Furthermore, the switching fields are symmetric about zero, indicating negligible exchange bias between the IrMn film and the thicker Co layer. 

The hysteresis loop due to current-driven reversal of the free-layer in an applied easy-axis field of 66.7 mT is shown in Fig. 1(b).  The resistance change between the two states is abrupt and similar in magnitude to what is seen in Fig. 1(a). The applied field opposes the effective field arising from the interlayer magnetostatic coupling and causes device bistability at zero current.  The current-dependent switching behavior is consistent with the prediction that the spin torque driven instability occurs for only one current direction for the given initial layer configuration\cite{myers2sw}.  Given the sign convention we have chosen for the current, theory predicts that positive currents stabilize the high-resistance, antiparallel state, whereas negative currents stabilize the low resistance, parallel state.  Accordingly, the switch to the high-resistance state occurs for positive current, whereas a negative current switches the device into the low resistance state.
%\begin{figure}
%\centerline{\includegraphics*[width=0.4\textwidth]{fig1c.eps}}
% \label{fig:mrloops}
%\end{figure}

To study device reversal driven by high-speed current pulses, a microwave probing assembly, schematically shown in Fig. 2, was used.  The device was contacted by coplanar waveguide (CPW) probes connected to a 50 $\Omega$ transmission line cable that connects to the 50 $\Omega$ input port of a bias-tee. The device resistance was monitored through the inductive path of the bias-tee while current pulses were applied through the capacitive path.  Several pulse generators were used, which allowed pulse duration to vary from 100 ps to 200 ns.  The pulse rise times ranged from about 50 ps for 100 ps to 10 ns pulses, 2 ns for 10 ns to 100 ns pulses, and 8 ns for longer pulses.  %Pulse amplitude was controlled using high bandwidth microwave attenuators.  

The current through the device for a given applied pulse was determined by considering the device as a resistive load $Z_L$ at the end of a 50 $\Omega$ transmission line (for the device considered here, $Z_L = 7.4\, \Omega$). The initial current flowing along the transmission line is $i = V_0/Z_0$, where $V_0$ is the pulse voltage and $Z_0 = 50\, \Omega$.  However, when the pulse reaches the device, power is reflected due to the impedance mismatch.  Thus, the current through the device is given by $i_L = (V_0/Z_0)(1-\Gamma)$, where $\Gamma = (Z_L - Z_0)/(Z_L + Z_0)$ is the reflection coefficient.

The measurement for determining switching probability due to current pulses follows a process similar to previous measurements made to determine the switching probability of current-in-plane spin valves due to applied field pulses\cite{russek_book2}.  A DC current of 3.5 mA was applied through the device for 1 s.  This was sufficient current to consistently set the device into the antiparallel state as seen in Fig. 1(b).  Next, a negative polarity current pulse was applied through the device.  The device resistance was measured and then reset again to the antiparallel state.  This process was repeated 100 times for identical pulses to give the switching probability for that particular pulse amplitude and duration. The switching probability was determined in this way for a systematically varied group of pulse durations and pulse amplitudes.  Throughout the measurement process, quasi-static, current-driven switching hysteresis loops were measured to determine whether the application of many current pulses altered the device properties:  The two-terminal device resistance changed slightly over time, but the resistance change between states and the switching currents did not vary.

\section{\label{sec:results}Results and Discussion}

The probability of switching from the antiparallel to the parallel state was measured as a function of the pulse duration for negative current pulse amplitudes increasing from -1.8 mA to -7.8 mA.  The quasi-static switching current, determined from Fig. 1(b), was -1.0 mA.   The pulse duration was varied in increments as small as 100 ps.  Figure 3 shows the switching probability as a function of pulse duration for several values of pulse amplitude.  The switching probability increases from zero to one as the pulse duration is increased.  This trend is consistent with two types of switching behavior: thermal activation over a barrier, and dynamic reversal, where the barrier between two states is completely suppressed.  In zero effective field and zero applied current, the device free-layer orientation is bistable, indicating an energy barrier separating two stable magnetization states.  The energy barrier between the states can be lowered by application of a magnetic field, leading to reversal and hysteresis. Since hysteretic reversal can also occur by way of applied currents, it is reasonable to propose that the mechanism of spin torque also serves to lower the energy barrier between the two magnetization states as current of the appropriate direction is increased through the device.  Evidence that spin torque can control the barrier height, and hence the thermal activation rate for switching, has been found in the results of a variety of experiments\cite{albert2sw, myers1sw, koch1swcpp, pufall_smt}. 

Thermally activated reversal is expected at lower values of current where the energy barrier separating the two stable states of the device is decreased during application of the pulse to a height of the order of $k_BT$, where $k_B$ is Boltzmann's constant and $T$ is room temperature for our measurements.  In this case, the time dependence of the probability for incurring a switch event is a simple exponential function\cite{brown_stochastic} $P(t_d) = 1 - e^{-t_d/\tau}$.  The pulse duration, $t_d$, is equivalently the time interval during which the energy barrier is lowered to allow for a thermally activated switching process.  The parameter $\tau$ is the average dwell time of the initial state before switching, and it depends on the ratio of the barrier height to the temperature, which would decrease with applied current.  At higher currents, where the energy barrier is completely suppressed, reversal takes place through a deterministic trajectory, and a simple exponential dependence cannot be expected.  Rather, the switching probability should remain zero until the pulse duration exceeds the period required for the magnetization to dynamically evolve past the device hard-axis direction (i.e., a state that relaxes to parallel alignment).  %This critical period %depends on the switching speed. 
Once the pulse duration exceeds this critical period, which depends on switching speed, the switching probability sharply increases to one.  The switching speed should increase with current\cite{sun_smt} which implies that for larger currents, smaller pulse durations are required for consistent switching. 

% At 300 K, there is some distribution of the initial magnetization direction about its equilibrium state which results in a %finite width in pulse duration over which the switching probability increases from zero to one (as seen in Fig. 3).  %Furthermore, this width increases as the current decreases indicating that the width is not simply due to fluctuations in the %pulse duration and amplitude from pulse to pulse. 

The switching probability vs. pulse duration data do not fit the simple exponential form predicted for thermal activation.  This disagreement is specifically shown in Figs. 4(a) and 4(b) for currents of -2.8 mA and -2.0 mA respectively.  For all the currents investigated, the exponential fits fail to match the sharp inflections in the data where the probability increases from zero and when the probability settles at one.  However, as the current amplitude decreases, the deviation from the exponential fit decreases, as seen in Fig. 4(b).  A much better fit to the data (also shown in Figs. 4(a) and 4(b)) is given by $1-g(t_d)$, where $g(t_d)= 1/(1+e^{(t_d-A)/B})$ is a Fermi distribution function with two free parameters A and B.

The data do show a critical duration that varies inversely with current, beyond which the switching probability becomes nonzero.  Furthermore, the switching probability changes smoothly, over a measurable width $t_w$, from zero to one.  For the case of switching while the barrier is fully suppressed, $t_w$ can be due to a distribution of initial magnetization directions about the equilibrium easy-axis direction at finite temperatures.  Thus, the response to identical pulses varies from pulse to pulse because differences in initial states will affect the switching speed and/or the size of the angle through which the magnetization needs to travel for reversal.  This effect produces a distribution of the critical pulse durations needed to reverse the device.  Our data show that the width of the distribution $t_w$ increases as the current decreases, further indicating that the width is not simply due to fluctuations in the pulse duration and amplitude from pulse to pulse.

For the case of switching while the barrier is completely suppressed, the pulse duration required for reversal is predicted to be inversely proportional to the current\cite{koch1swcpp,zhang_smt}.  
%In order to determine switching times, the switching probability vs. pulse duration data for all the current amplitudes measured, %were fit to a four parameter function of the form $P(t_d)=(A_1-A_2)/(1+e^{(t_d-A_3)/A_4})+A_2$, where the $A_{[n]}$ are the %fitting parameters.  This functional form provides a very close fit to the data, however no physical interpretation is currently %connected to this equation.  
From the switching probability vs. pulse duration data for all currents investigated, the point $\tau_{95}$, the duration at which the switching probability equals 0.95, is extracted. Figure 5 shows $1/\tau_{95}$ plotted vs. the current amplitude.    A linear relationship between $1/\tau_{95}$ and current is seen extending from -2 mA to -5.5 mA with a slope -0.55 GHz/mA.  The linear trend indicates driven dynamic reversal in the absence of a barrier, which has also been observed in micromagnetic simulations\cite{zhang_smt} and experimentally by a different measurement technique\cite{koch1swcpp}.  Beyond -5.5 mA, there is a kink and again a linear relationship with a slightly smaller slope.  For a current of -7.8 mA, $\tau_{95}$ is 290 ps.   Shorter switching times should be possible for these devices, as the devices are capable of withstanding DC currents up to -15 mA, and therefore larger pulsed-currents.  

For current amplitudes less than -2 mA, $1/\tau_{95}$ begins to curve toward the quasi-static switching current of -1 mA.  This is the region where a small energy barrier exists during the application of the current pulse, and thermally activated reversal occurs.  An extrapolation of the linear fit to the data between -2 mA and -5.5 mA gives a value for the zero-temperature switching current $I_{c0} = (-1.91 \pm 0.08)$ mA.  The switching probability measured at -1.8 mA still does not fit an exponential form very well, indicating that this current puts the device in a complicated transition region between fully dynamic and thermally activated reversal.  It is expected that the measured switching probability vs. pulse duration for currents between -1.8 mA and -1 mA will fit an exponential dependence.  Previous work has shown exponentially dependent switching probabilities with barrier heights that depend on current\cite{koch1swcpp,albert2sw, myers1sw}.  However, the behavior at currents close to the quasi-static switching current cannot be fully investigated with the present experimental setup because the long-duration pulses ($>500$ ns) required to observe exponential behavior are severely distorted by the capacitance of the bias-tee.  

% over a large range of pulse durations, as predicted for thermally activated processes.  

In summary, pulsed-current switching due to spin momentum transfer occurs in CPP spin valve nanopillars for pulses as short as 290 ps.  Based on the measured relationship between the threshold pulse duration required for consistent switching and the pulse amplitude, the transition between thermally activated reversal and fully dynamic reversal was determined.  The current at this transition should be the zero-temperature switching current for the device for the particular applied field used.

%  data point at 1.73 mA - not quite a good exponential fit - indicating a need to take more data in that region of low currents,
%region really a transition between thermal activation and dynamics

%point - do not see switching to partial switched states as in CIP device.  Could it be that the devices are more single domain, %or is it that we cant really measure it here - what is the noise level in the measurement of resistance.  In CIP devices it was typically 0.01 ohm i think, here we are looking at 0.2 ohm full changes.

% Essentially, the device is reset to the high-resistance state by the application of -xxx mA of current for 1 second along the DC %path.

%Device response both to field and current fig1

%device in meas apparatus fig 2

%In order to produce current densities of the order $10^7$ A/cm$^2$ required to achieve current induced switching, small dimension %devices are 

\bibliography{cppswbib,refs2}% Produces the bibliography via BibTeX.
\pagebreak
\section{\label{sec:captions}Figure Captions}
\newcounter{hello}
\begin{list}
{Fig. \arabic{hello}}{\usecounter{hello} \setlength{\rightmargin}{\leftmargin}}

\item (a)  $dV/dI$ vs. easy-axis magnetic field for 50 nm $\times$ 100 nm hexagon shaped CPP spin valve nanopillar.  The dashed line indicates the response as field is positively ramped, and the dotted line is the response to negative field ramp.  (b) $dV/dI$ vs. current showing current-driven reversal in an applied field $\mu_0$H = 66.7 mT.  Arrows indicate the polarity of the resistance change.

\item  Schematic of apparatus used to measure pulsed-current response of the CPP spin valve nanopillars.  The center conductor of the CPW probe contacts the top lead of the device, while the other two ground pins contact the bottom of the device.  Double lines indicate 50 $\Omega$ transmission lines.

\item  Switching probability as a function of pulse duration for 5 values of current amplitude.  Squares denote the response to -7.8 mA pulses, circles for -5.5 mA, upward triangles for -3.9 mA, downward triangles for -2.8 mA, and diamonds for -2 mA.

\item  Comparison of switching probability data (squares) with a simple exponential fit (solid line) and a better fit using the Fermi distribution function (dashed line). (a) Data and fits for $i_L = -2.8$ mA. (b) Data and fits for $i_L = -2.0$ mA. 

\item  Reciprocal of the pulse duration required to achieve 95\,\% switching probability, $1/\tau_{95}$, for all current amplitudes that were investigated.  $\tau_{95}$ was determined from a data set including those data shown in Fig. 3.  The solid line is the fit to the data between -2 mA and -5.5 mA.

\end{list}

\pagebreak
\begin{figure}
\centerline{\includegraphics*[width=\textwidth]{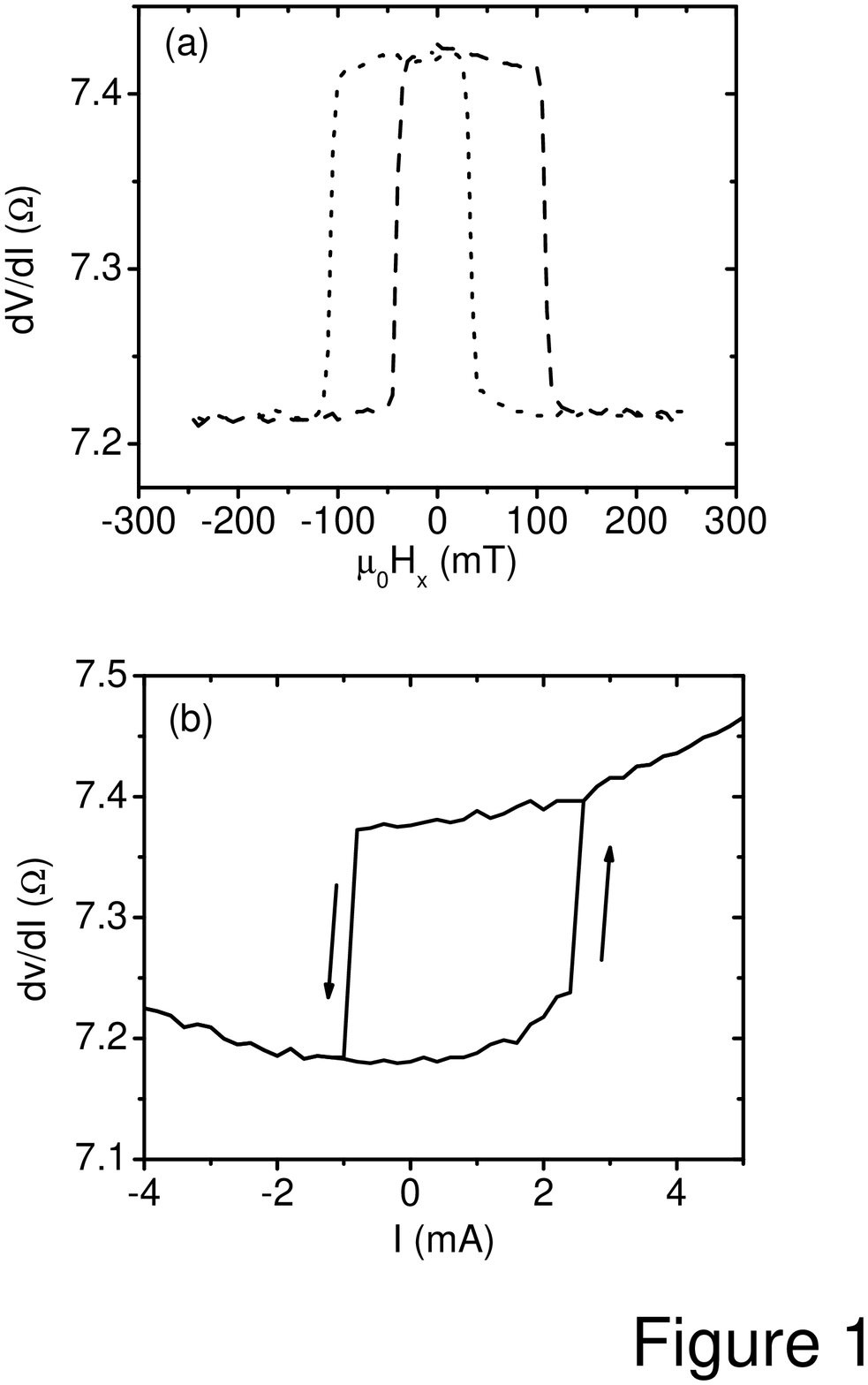}}
 \label{fig:mrloops}
\end{figure}

\pagebreak

\begin{figure}
\centerline{\includegraphics*[width=\textwidth]{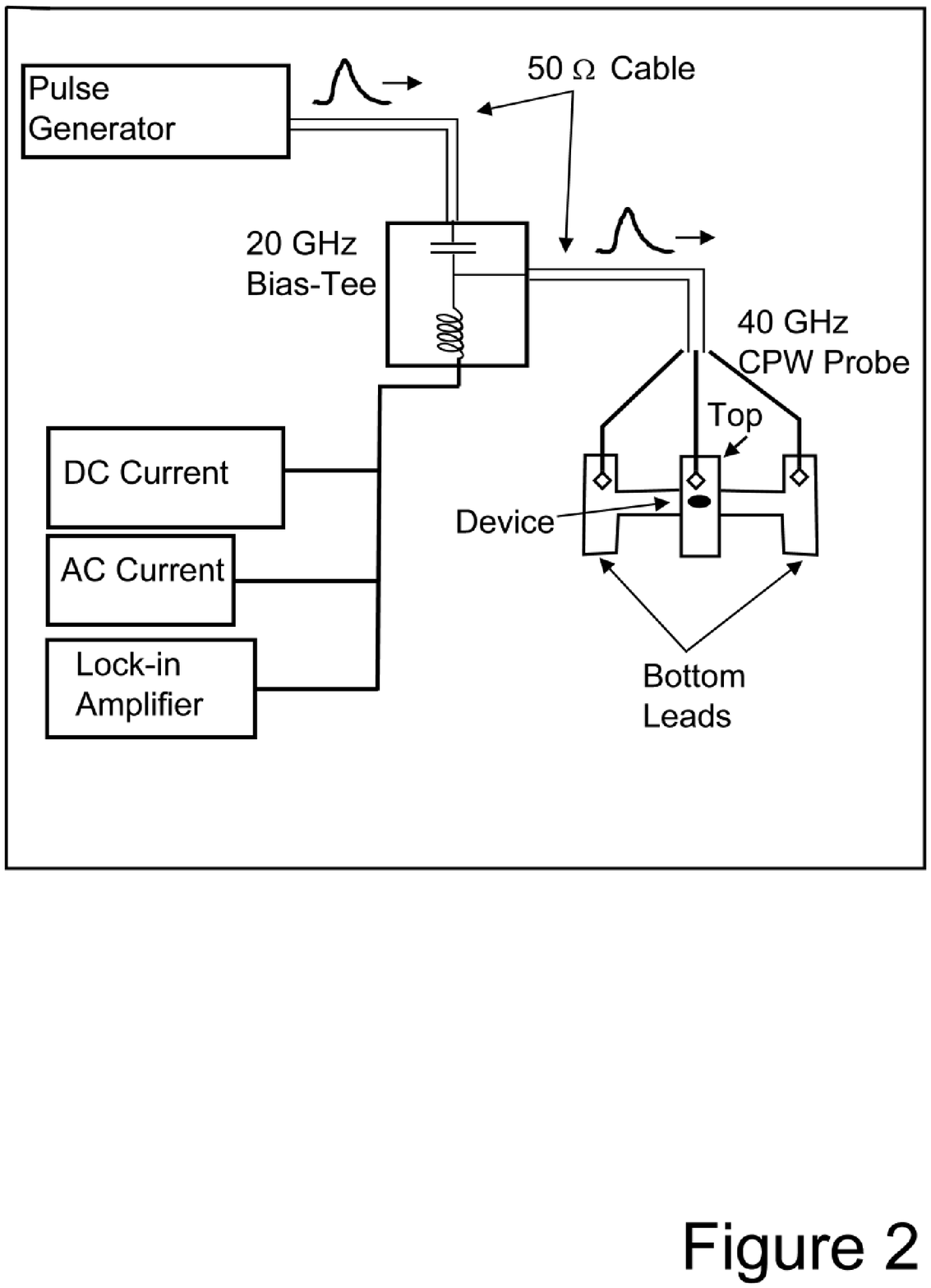}}
 \label{fig:schematic}
\end{figure}

\pagebreak
\begin{figure}
\centerline{\includegraphics*[width=\textwidth]{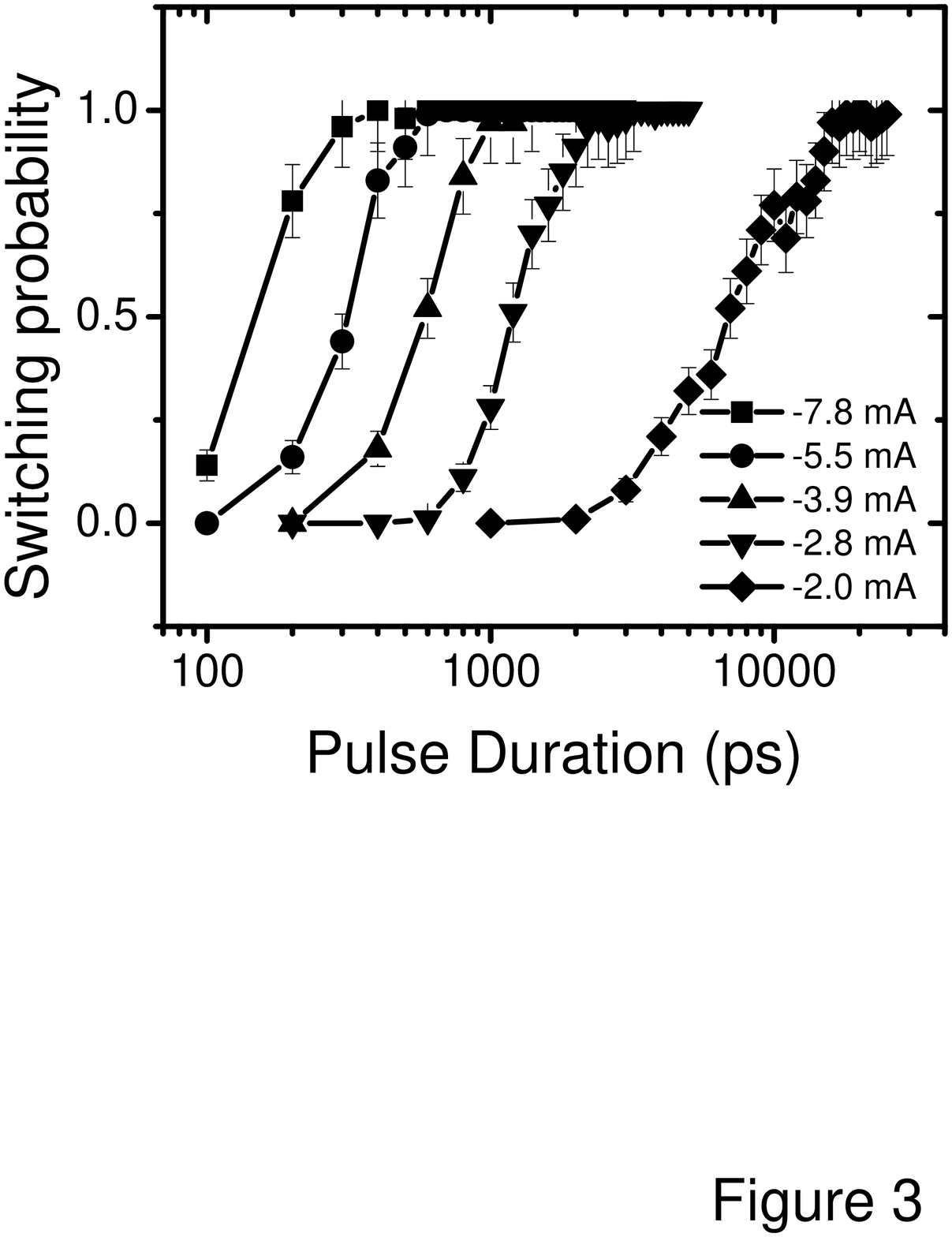}}
 \label{fig:swprob}
\end{figure}

\pagebreak
\begin{figure}
\centerline{\includegraphics*[width=\textwidth]{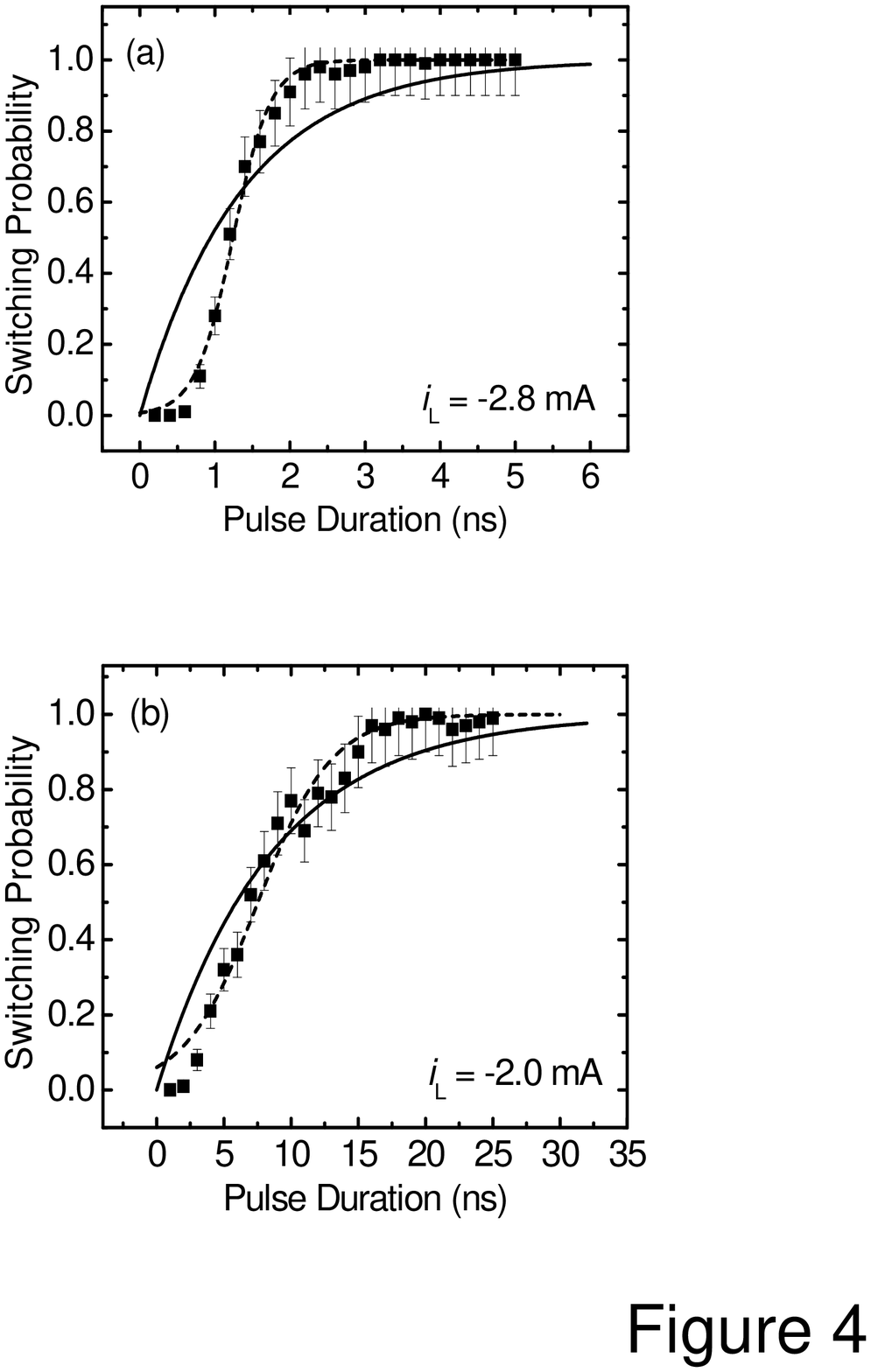}}
 \label{fig:tau}
\end{figure}

\pagebreak
\begin{figure}
\centerline{\includegraphics*[width=\textwidth]{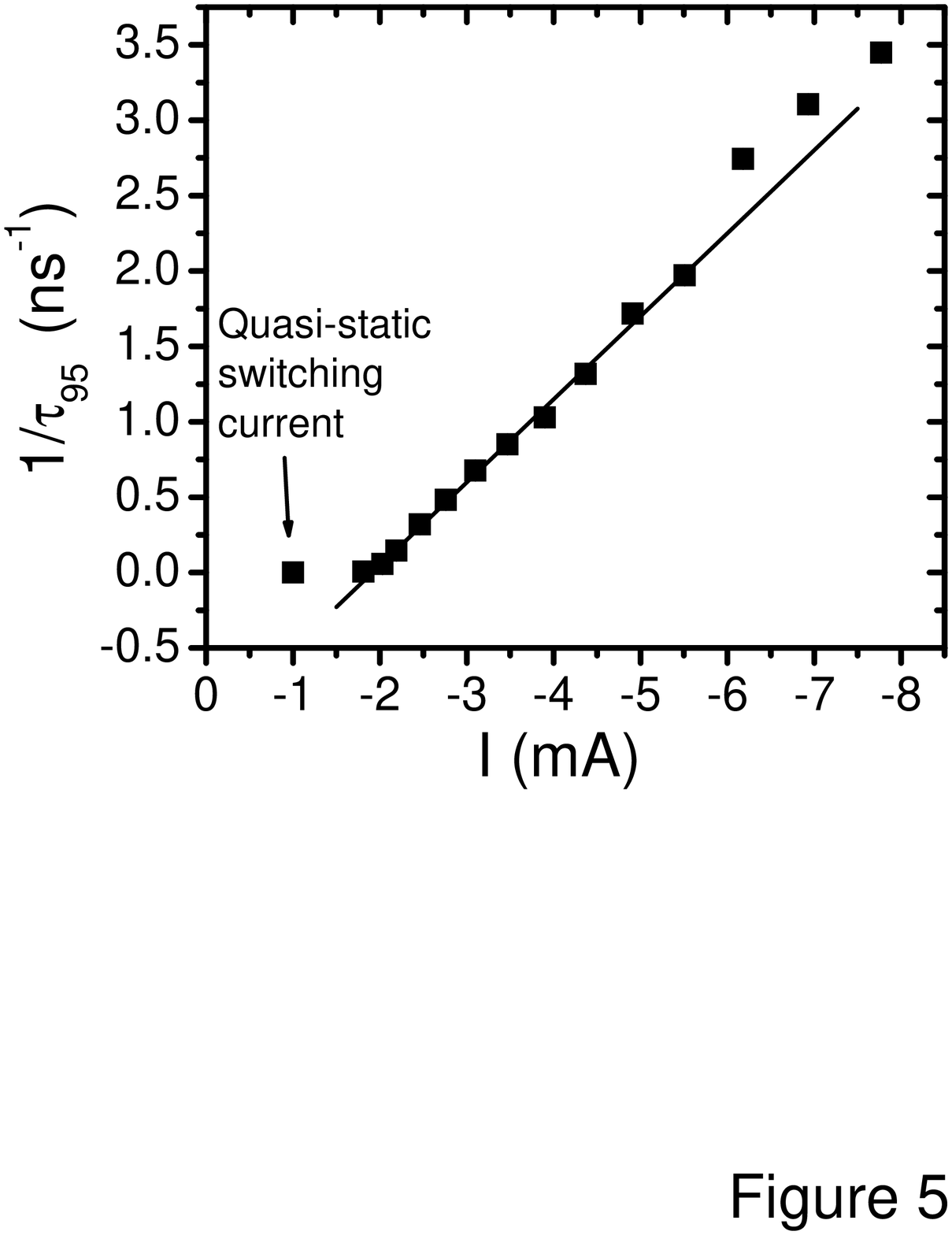}}
 \label{fig:tau}
\end{figure}

\end{document}